\documentclass[twocolumn, prl, superscriptaddress,notitlepage]{revtex4}
\usepackage{graphicx}
\usepackage{physics}
\usepackage{epsfig}
\usepackage{color}
\usepackage{xspace}
\usepackage{array}
\usepackage{amsmath}
\usepackage{amsfonts}
\usepackage{ulem}
\usepackage[dvipsnames]{xcolor}
\usepackage{colortbl}
\usepackage[breaklinks,colorlinks,linkcolor=blue,citecolor=blue,urlcolor=blue]{hyperref}

\begin{document}
\title{Universal Hypothesis of Autocorrelation Function from Krylov Complexity}

\author{Ren Zhang}
\affiliation{School of Physics, Xi'an Jiaotong University, Xi'an, Shaanxi 710049, China }
\author{Hui Zhai}
\email{ hzhai@tsinghua.edu.cn}
\affiliation{Institute for Advanced Study, Tsinghua University, Beijing, 100084, China}

\date{\today}

\begin{abstract}
In a quantum many-body system, autocorrelation functions can determine linear responses nearby equilibrium and quantum dynamics far from equilibrium. In this letter, we bring out the connection between the operator complexity and the autocorrelation function. In particular, we focus on a particular kind of operator complexity called the Krylov complexity. We find that a set of Lanczos coefficients $\{b_n\}$ computed for determining the Krylov complexity can reveal the universal behaviors of autocorrelations, which are otherwise impossible. When the time axis is scaled by $b_1$, different autocorrelation functions obey a universal function form at short time. We further propose a characteristic parameter deduced from $\{b_n\}$ that can largely determine the behavior of autocorrelations at the intermediate time. This parameter can also largely determine whether the autocorrelation function oscillates or monotonically decays in time. We present numerical evidences and physical intuitions to support these universal hypotheses of autocorrelations. We emphasize that these universal behaviors are held across different operators and different physical systems. 
\end{abstract}

\maketitle

The autocorrelation function refers to the temporal correlation of the same operators $\hat{O}$ at two different times, denoted by $\mathcal{C}(t)=\langle \hat{O}(t)\hat{O}(0)\rangle$. It plays an important role in studying quantum matters, including condensed matter materials, ultracold atomic gases, and NMR systems. The autocorrelation functions can be measured by spectroscopy methods, which reveals linear response nearby equilibrium. It can also be measured in far-from-equilibrium dynamics, for example, through quench experiments. In non-equilibrium situations, autocorrelation functions can determine quantum dynamics involving highly excited states. 

The behavior of the autocorrelation function $\mathcal{C}(t)$ crucially depends on the time evolution of $\hat{O}(t)$, which follows the Heisenberg evolution $\hat{O}(t)=e^{i\hat{H}t}\hat{O}e^{-i\hat{H}t}$, with $\hat{H}$ being the Hamiltonian of the physical system. Usually, $\hat{O}(t)$ becomes more and more complicated as time evolves, consequently reducing the autocorrelation $\mathcal{C}(t)$. This observation brings out the connection between operator complexity and its autocorrelation function. In the past years, various measures have been proposed to describe operator complexity quantitatively and to study how the complexity of an operator grows under the Heisenberg evolution \cite{B. Yoshida, R.C. Myers, A. Streicher, R.-Q. Yang, S. Sharma, K.-Y. Kim, A. Streicher-2, A.Lucas, O. Parrikar, O. Parrikar-2, Altman}. In particular, the Krylov complexity has been proposed to quantify operator complexity recently~\cite{Altman}. The advantage of the Krylov complexity is that it exhibits universal behavior for generic chaotic quantum many-body systems, and such universal behaviors are shared by a large class of different quantum many-body Hamiltonians~\cite{R. Sinha,A. Dymarsky,A. Gorsky-2,Z. Y. Xian,J. Sonner,M. Smolkin,J. D. Noh,C. J. Lin,D. Shouvik,D. Patramanis,D. Patramanis-2,Chenwei,J. Sully,D. Rosa,A. del Campo,J. Sonner-2,T. Pathak,Q. Wu,J. Gemmer,S. Choudhury,A. Roy,S. Liu,Y. Yang,S. Pawar,Z. Y. Fan,Z. Y. Fan-2,J. Sonner-3,Zhai,Jian}. 

Therefore, it is natural to ask whether these recent developments in measuring operator complexity can help us better understand the autocorrelation function's universal behavior. Especially, since the Krylov complexity exhibits universality, the question is, by utilizing the notation introduced for studying the Krylov complexity, whether we can reveal hidden universal behaviors of the autocorrelation function, which are otherwise impossible.  

Let us first introduce the notation for describing the Krylov complexity \cite{Altman}. Given a Hilbert space spanned by $\{\ket{i}\}$, an operator $\hat{O}=\sum_{ij}O_{ij}\ket{i}\bra{j}$ can be mapped to a state $\ket{\hat{O}}$ in the double space as  $\ket{\hat{O}}=\sum_{ij}O_{ij}\ket{i}\otimes\ket{j}$, and vice versa. By using the Baker-Campbell-Hausdorff formula, the Heisenberg evolution can be expressed as 
\begin{equation}
\ket{\hat{ O}(t)}=\sum_{n}\frac{(it)^n}{n!}\ket{\hat{L}^n\hat{O}}, \label{BCH}
\end{equation}
where $\hat{L}\hat{O}\equiv [\hat{H},\hat{O}]$. Eq. (\ref{BCH}) can be viewed as expanding a state $\ket{\hat{O}}$ under a set of states $\{\ket{\hat{L}^n\hat{O}}\}$. However, this set of states are neither orthogonal nor normalized. Applying the Gram-Schmidt procedure to this set of states yield a set of orthogonal basis, denoted by the Krylov basis $\ket{\hat{\mathcal{W}}_n}$~\cite{Altman}. Here the inner product is introduced as $\bra{\hat{O}_1}\ket{\hat{O}_2}=\text{Tr}[\hat{O}^\dag_1\hat{O}_2]$. In general, the larger $n$, the more complicated the operator $\hat{\mathcal{W}}_n$. Expanding $\ket{\hat{O}(t)}$ under the Krylov basis gives 
\begin{equation}
\ket{\hat{O}(t)}=\sum_{n}\varphi_n (t) \ket{\hat{\mathcal{W}}_n}.
\end{equation}
It has been shown that $\varphi_n(t)$ obeys the following differential equation 
\begin{equation}
i\partial_t\varphi_n(t)=-b_{n+1}\varphi_{n+1}(t)-b_n\varphi_{n-1}(t).  \label{half-infinite-chain}
\end{equation}
The Krylov complexity is defined as $K(t)=\sum_n n|\varphi_n(t)|^2$. 
Here $\{b_n\}$ are called the Lanczos coefficients introduced in the Gram-Schmidt procedure. They depend on both the choice of operator $\hat{O}$ and the system Hamiltonian $\hat{H}$. The units of $\{b_n\}$ are energy. Note that $\ket{\hat{\mathcal{W}}_0}=\frac{1}{b_0}\ket{\hat{O}}$ with a normalization factor $b^2_0=\bra{\hat{O}}\ket{\hat{O}}$, we have $\varphi_0=\bra{\hat{\mathcal{W}}_0}\ket{\hat{O}}=\frac{1}{b_0}\mathcal{C}(t)$. That is to say, the time evolution of $\varphi_0$ gives rise to the autocorrelation function, normalized by its value at $t=0$.  

Eq. (\ref{half-infinite-chain}) is reminiscent of the Schr\"odinger equation for a single particle hopping along a half-infinite chain. At $t=0$, only $\varphi_0=1$ and all other $\varphi_{n\neq 0}=0$. As time evolves, this particle hops away from $n=0$, consequently reducing the autocorrelation $\mathcal{C}(t)$ and increasing the Krylov complexity. Inspired by this connection, it is realized that the dynamics of $\mathcal{C}(t)$ is governed by the Lanczos coefficients $\{b_n\}$. Hence, if there exist certain universal behaviors of $\mathcal{C}(t)$, it is conceivable that their information is hidden inside $\{b_n\}$. 

\textit{Summary of the Main Results.} Before presenting the details, we first summary the main findings of this work. First of all, we note that originally the Lanczos coefficients $b_n$ are only defined for all non-negative integers. However, we can use these data, especially the data with small $n$, to interpolate a smooth function $b[x]$ defined for all $x>0$. We require that the interpolated function has to be smooth enough and cannot strongly vary between two neighboring integers. Such an interpolation allows us to take derivatives of $b[x]$ and the first-order derivative is denoted by $b^\prime[x]$. With the help of $b[x]$ and $b^\prime[x]$, we present the following three universal hypothesis of $\mathcal{C}(t)$. 

1. \textbf{Short-Time Universality:} When the time $t$ is scaled by $b[1]$, the autocorrelation function ${\cal C}$ as a function of $t b[1]$ exhibits universal behavior for $tb[1]\lesssim 1$ across all different choices of operators and Hamiltonians. 

2. \textbf{Intermediate-Time Universality:} If two different operators with two different Hamiltonians share the same value of $b^\prime[1]/b[1]$, the autocorrelation function ${\cal C}(tb[1])$ exhibits universal behavior for $tb[1]\sim \mathcal{O}(1)$.

3. \textbf{Oscillatory Behavior:} There exists a general trend that ${\cal C}(tb[1])$ exhibits an oscillatory behavior when when $b^\prime[1]/b[1]$ is small, and ${\cal C}(tb[1])$ monotonically decays when $b^\prime[1]/b[1]$ is larger. 

We have run extensive numerical tests which support these hypothesis. Below, we first present a set of representative numerical evidences, and the complete code to verify our hypothesis are also available \cite{code}.

\begin{figure}[t]
    \centering
    \includegraphics[width=0.48\textwidth]{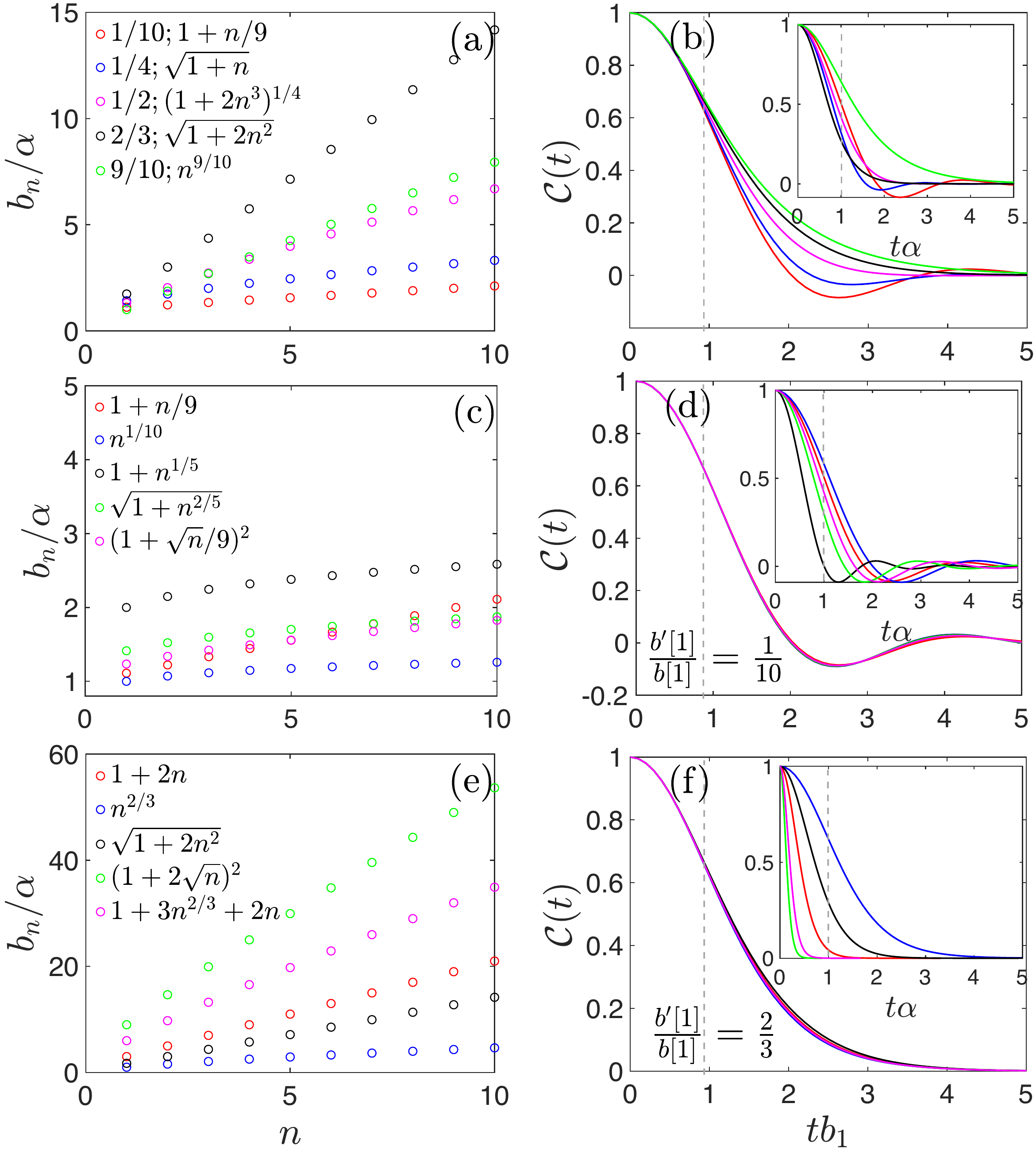}
    \caption{(a),(c) and (e) show several different choices of $b_n$ generated by $b[n]=\alpha f(n)$, with different functions $f(n)$ shown in the legend. In (a), $b^\prime[1]/b[1]$ take five different values as shown by the numbers in the legend. $b^\prime[1]/b[1]=1/10$ in (b) and $b^\prime[1]/b[1]=2/3$ in (c) for all choices of different functions of $b[x]$. (b), (d) and (f) show the autocorrelation function $\mathcal{C}(t)$ as a function of $tb_1$, corresponding to $b_n$ shown in (a), (c) and (e). The insets in (b), (d) and (f) show $\mathcal{C}(t)$ as a function of $t\alpha$.}
     \label{Gedanken}
\end{figure}

\textit{Gedanken Simulations.} Here we first present a Gedanken numerical simulation. This simulation does not involve physical Hamiltonians and only solves Eq. (\ref{half-infinite-chain}) with a set of given $\{b_n\}$. We assume $b[n]=\alpha f(n)$, where $\alpha$ is an energy unit and $f(n)$ can choose different smooth functions. The general form of $f(n)$ is $(a+b n^\beta)^\gamma$. By choosing different values of $a$, $b$, $\beta$ and $\gamma$, $f(n)$ can realize different functions as shown in the inset of Fig. \ref{Gedanken}. Each function generates a set of $\{b_n\}$ at integer values of $n$, and by solving Eq. (\ref{half-infinite-chain}) with this set of $\{b_n\}$, we can obtain $\varphi_0(t)$ as normalized $\mathcal{C}(t)$. The results are shown in Fig. \ref{Gedanken}. 

As shown in the inset of Fig. \ref{Gedanken}, when $\mathcal{C}(t)$ is plotted in terms of $t\alpha$, $\mathcal{C}(t)$ deviates from each other once $t\alpha$ is non-zero. However, when $\mathcal{C}(t)$ is plotted in terms of $tb_1$, all $\mathcal{C}(tb_1)$ collapse when $tb_1\lesssim 1$. This difference can be seen clearly by comparing Fig. \ref{Gedanken}(b) with its inset. This demonstrates the first hypothesis. 

However, these curves shown in Fig.  \ref{Gedanken}(b) do not collapse when $tb_1\gtrsim 1$. To exhibit universality in the intermediate time, we require that different set of $\{b_n\}$ share a common characteristic parameter. A main finding of this work is this characteristic parameter, which turns out to be $b^\prime[1]/b[1]$. In Fig. \ref{Gedanken}(c) and (e), we show two sets of $\{b_n\}$. Within each set, the values of $\{b_n\}$ are very different. However, they share the same value of $b^\prime[1]/b[1]$. It is easy to check that the five different sets of $\{b_n\}$ shown in Fig. \ref{Gedanken}(c) and Fig. \ref{Gedanken}(e) respectively share $b^\prime[1]/b[1]=1/10$ and $b^\prime[1]/b[1]=2/3$.  We then calculate $\mathcal{C}(t)$ using $\{b_n\}$ shown in Fig. \ref{Gedanken}(c) and (e), which respectively lead to results shown in Fig. \ref{Gedanken}(d) and (f). The inset of Fig. \ref{Gedanken}(d) and (f) also show that these autocorrelations $\mathcal{C}(t)$ behave very different in term of $t\alpha$. However, when they are plotted in terms of $tb_1$, Fig. \ref{Gedanken}(d) and (f) show that they perfectly collapse into a single curve up to $tb_1 \sim 5$. This demonstrates the second hypothesis. 

The numbers shown in the legend of Fig. \ref{Gedanken}(a) are $b^\prime[1]/b[1]$ for different choices of $\{b_n\}$. We find that when $b^\prime[1]/b[1]< \sim 0.5$, the corresponding $\mathcal{C}(t)$ all exhibit oscillatory behavior. This can also be seen from all cases in Fig. \ref{Gedanken}(c) and (d). When $b^\prime[1]/b[1]> \sim 0.5$, the corresponding $\mathcal{C}(t)$ all monotonically decay, as one can also seen from all cases in Fig. \ref{Gedanken}(e) and (f). This demonstrates the third hypothesis. All together, 
the Gedanken simulation shows
that the autocorrelation function exhibits universal behavior in terms of $tb_1$ and a single parameter $b^\prime[1]/b[1]$ can largely determine the behavior of $\mathcal{C}(t)$.  

\textit{Physical Model.} The Gedanken numerical simulation is inspiring and generic because it does not depend on the concrete physical operator and physical Hamiltonian. However, it also has limitations because the Lanczos coefficients $\{b_n\}$ obtained from physical models are usually not smooth enough. To show how much the discussion above can hold for realistic models, we show results calculated with two typical physical models below. 

The first model is the one-dimensional quantum Ising model with both transversal and longitudinal fields, whose Hamiltonian is written as  
\begin{equation}
\hat{H}_\text{Ising}=\sum_{i}\left(J\hat{\sigma}^z_i\hat{\sigma}^z_{i+1}+h\hat{\sigma}^x_i+g\hat{\sigma}^z_i\right).
\end{equation} 
$h/J$ and $g/J$ are two tunable parameters in this model. The second model is the one-dimensional spinless fermion Hubbard model, whose Hamiltonian is written as  
\begin{align}
\hat{H}_\text{Hubbard}=-\sum_{i}&(J\hat{c}^\dag_i\hat{c}_{i+1}+J^\prime\hat{c}^\dag_i\hat{c}_{i+2}+\text{h.c.}\nonumber\\
&+V_1\hat{n}_i\hat{n}_{i+1}+V_2\hat{n}_i\hat{n}_{i+2}).
\end{align}
Here $J$ and $J^\prime$ denote the nearest and the next nearest neighbor hopping strengths. $V_1$ and $V_2$ are the nearest and the next nearest interaction strengths. In both models, we use $J$ as the natural energy unit although they represent different energy scales in two models.  

\begin{figure}[t]
    \centering
    \includegraphics[width=0.47\textwidth]{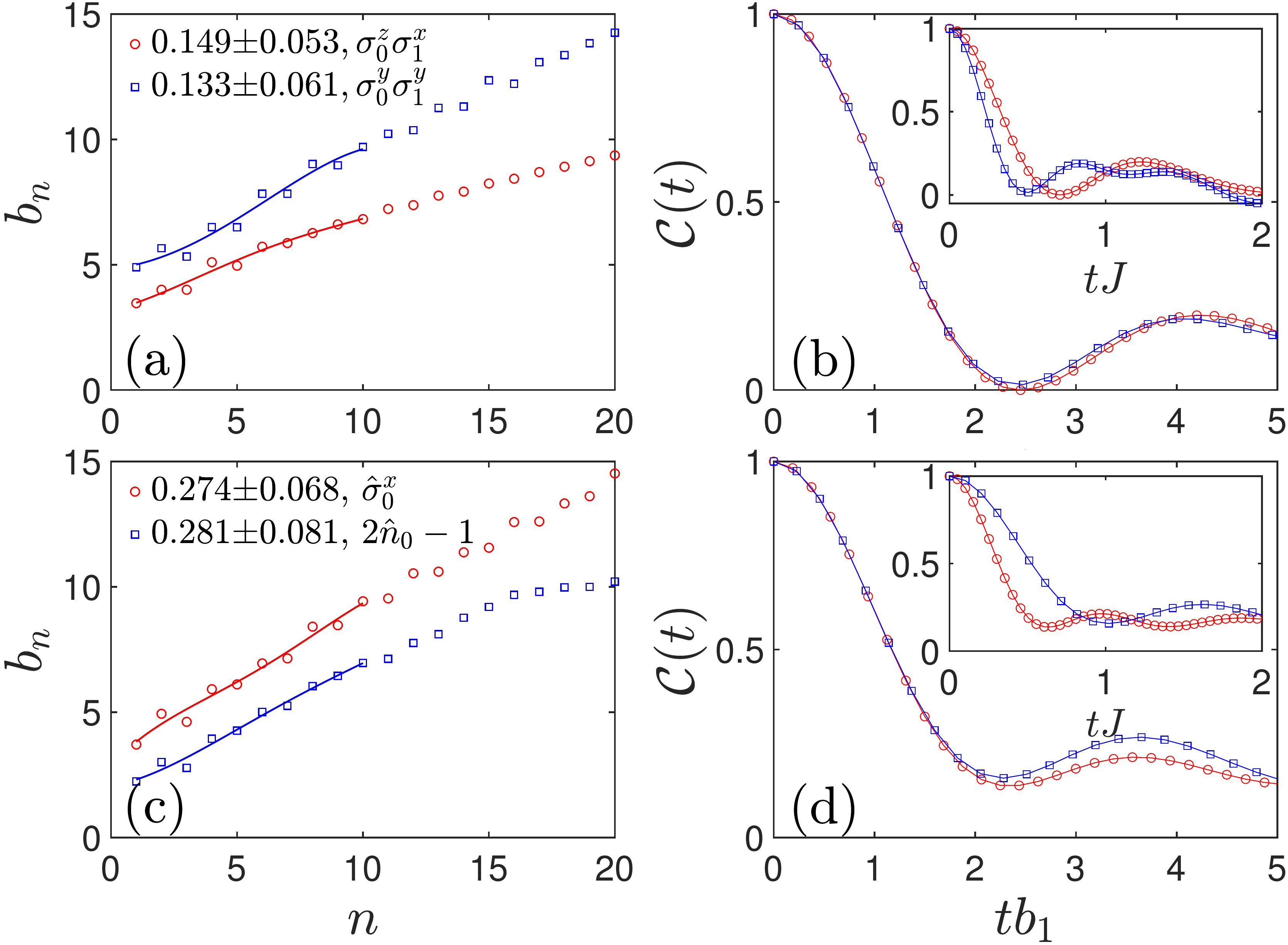}
    \caption{(a-b) The Lanczos coefficients $b_n$ and the autocorrelation functions $\mathcal{C}(t)$ for two different operators $\hat{\sigma}^z_0\hat{\sigma}^x_1$ (blue squares) and $\hat{\sigma}^y_0\hat{\sigma}^y_0$ (red circles) in the quantum Ising model. $(h/J,g/J)=(1,0)$ for blue squares and $=(1,1)$ for red circles. (c-d) $b_n$ and $\mathcal{C}(t)$ for two different operators in two different systems. Red circles are results for $\hat{\sigma}^x_0$ in the quantum Ising model with $(h/J,g/J)=(1,1.2)$, and blue squares are results for $2\hat{n}_0-1$ in the spinlees Hubbard model with $(J^\prime/J,V_1/J,V_2/J)=(0.5,2,0.5)$. Here we calculate $13$ sites with periodic boundary condition for the quantum Ising model and $14$ sites for the spinless Hubbard model, and the lower index $0$ and $1$ represent site indices. In (b) and (d), $\mathcal{C}(t)$ is plotted in terms of $tb_1$ and is plotted in terms of $tJ$ in its inset. The sold lines in (a) and (c) are fitting curve of $b[x]$, and the numbers in the legend of (a) and (c) show $b^\prime[1]/b[1]$ for different cases, with error bars from fitting error.    }
     \label{universal}
\end{figure}

In Fig. \ref{universal}, we first compare $\mathcal{C}(t)$ of two different operators in the quantum Ising model, but with different model parameters, as shown in Fig. \ref{universal}(a) and (b). Then, we compare $\mathcal{C}(t)$ for two different operators in two different physical systems, one in the quantum Ising model and the other in the spinless Hubbard model. The results are shown in Fig. \ref{universal}(c) and (d). The Lanczos coefficients $\{b_n\}$ are now calculated with the chosen operators and the physical Hamiltonian. Then we use polynomial function to fit $\{b_n\}$ with respect to a few smallest integer $n$, up to $n\sim 10$. With the fitted function, we can obtain $b^\prime[1]/b[1]$ with an error bar from the fitting error. The fitting error also includes fluctuation due to varying the range of polynomial function and varying the number of integer points included in the fitting. 

Each figure of Fig. \ref{universal}(a) and (c) shows two cases with same value of $b^\prime[1]/b[1]$ within the error bars. Their corresponding $\mathcal{C}(t)$ in terms of $tb_{1}$ are respectively shown in Fig. \ref{universal}(b) and (d), compared with $\mathcal{C}(t)$ plotted in terms of $tJ$ in the insets. It is clear that by changing $tJ$ to $tb_1$, the horizontal axes is stretched such that the two cases shown in each figure are in good agreement with each other up to $tb_1\sim 2-5$. 

\begin{figure}[t]
    \centering
    \includegraphics[width=0.47\textwidth]{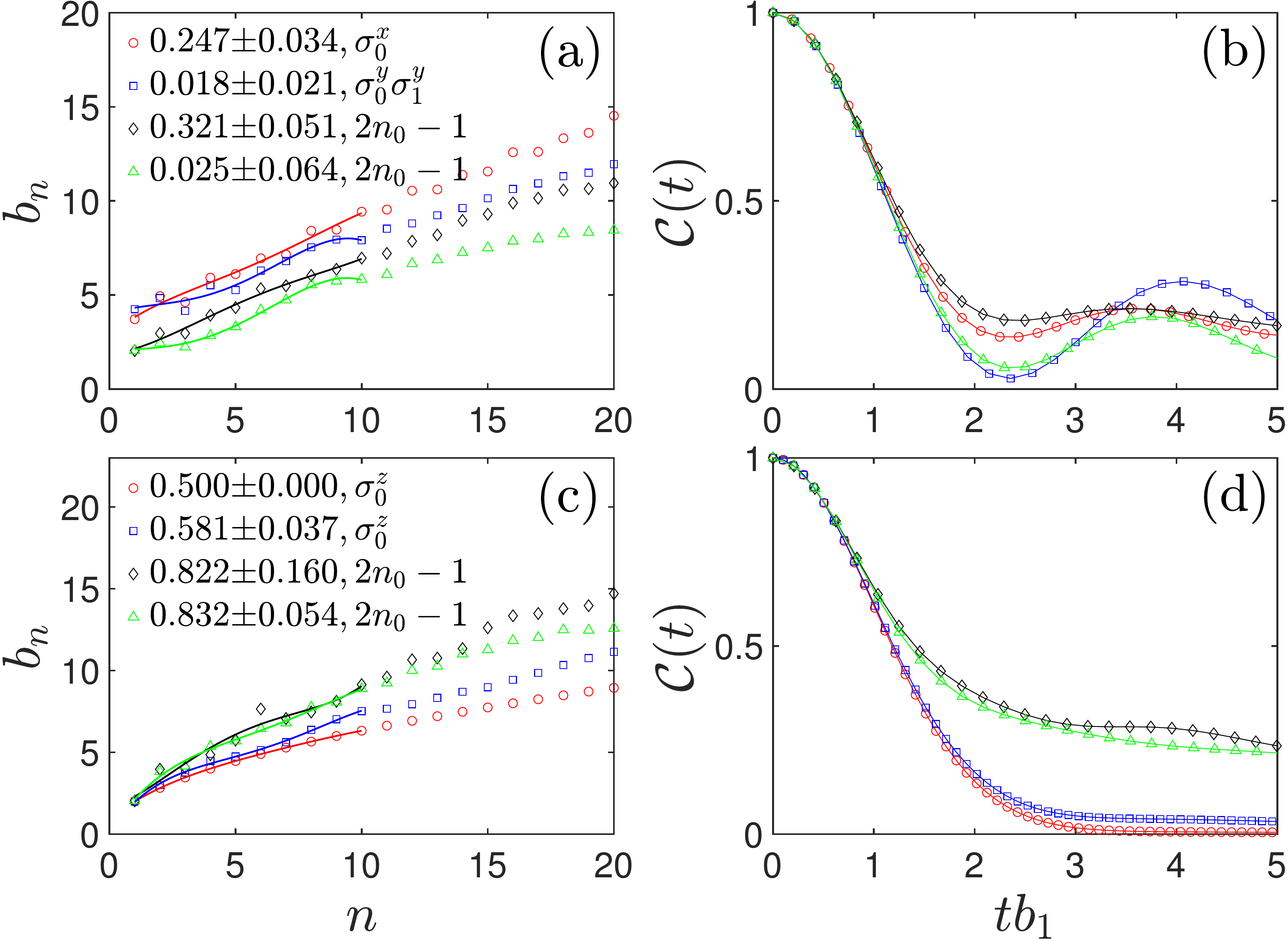}
    \caption{(a) The Lanczos coefficients $b_n$ for two operators $\hat{\sigma}^x_0$ (red circles) and $\hat{\sigma}^y_0\hat{\sigma}^y_1$ (blue squares) in the quantum Ising model, and the operator $2\hat{n}_0-1$ for the spinless Hubbard model (black diamonds and green triangles). The model parameters are $(h/J,g/J)=(1,1.2)$ for red circles and $=(1,0.5)$ blue squares, and $(J^\prime/J,V_1/J,V_2/J)=(0.2,3,0.5)$ for black diamonds and $=(0.21,1,1)$ for green triangles. The corresponding $\mathcal{C}(t)$ is plotted in terms of $tb_1$ in (b). (c) $b_n$ for the operators $\hat{\sigma}^z_0$ (red circles and blue squares) in the quantum Ising model, and the operator $2\hat{n}_0-1$ for the spinless Hubbard model (black diamonds and green triangles). The model parameters are $(h/J,g/J)=(1,0)$ for red circles and $=(1,0.5)$ blue squares, and $(J^\prime/J,V_1/J,V_2/J)=(0.2,5,0.5)$ for black diamonds and $=(0.21,1,3.5)$ for green triangles. The corresponding $\mathcal{C}(t)$ is plotted in terms of $tb_1$ in (d). Other information about the system size, the solid fitting lines and the numbers in the legend are the same as described in the caption of Fig. \ref{universal}.   }
     \label{oscillation}
\end{figure}

Fig. \ref{oscillation}(a) shows four cases with small $b^\prime[1]/b[1]$, with two from the quantum Ising model and two from the spinless Hubbard model. Their corresponding $\mathcal{C}(t)$ all exhibit the oscillatory behavior. On contrast, Fig. \ref{oscillation}(c) show four cases from these two models with larger $b^\prime[1]/b[1]$. Their corresponding $\mathcal{C}(t)$ all monotonically decay. Moreover, we note that because all these cases have different values of $b^\prime[1]/b[1]$, therefore their $\mathcal{C}(t)$ do not collapse when $tb_1\gtrsim 1$ but they all collapse when $tb_1\lesssim 1$. Hence, the information presented in Fig. \ref{universal} and Fig. \ref{oscillation} demonstrate the three hypothesis in realistic models. 

\textit{Intuitions.} Finally we discuss the intuitions that lead to these hypothesis. First, since all $\varphi_n=0$ at $t=0$ except for $\varphi_0$, and $\varphi_0$ only couples to $\varphi_1$, the short-time dynamics of $\varphi_0$ is dominated by its coupling to $\varphi_1$. Hence, by ignoring all $\varphi_{n>1}$, we can obtain 
\begin{equation}
i\partial_t\varphi_0=-b_1\varphi_1, \    \ i\partial_t\varphi_1=-b_1\varphi_0.
\end{equation} 
This gives rise to a solution $\varphi_0=\cos(tb_1)$. This shows that $b_1$ is a natural unit to scale $t$, resulting in a universal function form for autocorrelation functions at short-time. 

Secondly, we take the continuum limit of Eq. (\ref{half-infinite-chain}), which results in the following differential equation 
\begin{equation}
i\partial_t\varphi=-2(b[x]+b^\prime[x])\varphi-b^\prime[x]\partial_x\varphi-b[x]\partial^2_x\varphi. 
\end{equation}
Now if we make a frame transformation to redefine $t^\prime=tb[x]$, we arrive at the following equation
\begin{equation}
i\partial_{t^\prime}\varphi=-2\left(1+\frac{b^\prime[x]}{b[x]}\right)\varphi-\frac{b^\prime[x]}{b[x]}\tilde{\partial}\varphi-\tilde{\partial}^2\varphi,
\end{equation}
where $\tilde{\partial}$ is defined as
\begin{equation}
\tilde{\partial}=t^\prime \left(\frac{b^\prime[x]}{b[x]}\right)\partial_{t^\prime}+\partial_x.
\end{equation}
Hence, in the new frame, the differential equation is solely controlled by $b^\prime[x]/b[x]$. Since the autocorrelation function only concerns the time dynamics of $\varphi$ at $x=0$, we conjecture that $b^\prime[1]/b[1]$ largely determines the behavior of the autocorrelation.   

Thirdly, as for why $b^\prime[1]/b[1]\sim 0.5$ separates oscillating decay and monotonically decay, the intuition comes from solving simple situation with $b_n=\alpha n^\delta$. In these situations, $b[1]=\alpha$ and $b^\prime[1]/b[1]=\delta$. It turns out that Eq. (\ref{half-infinite-chain}) can be solved exactly for $\delta=0$, $1/2$ and $1$, which respectively give
\begin{align}
\label{Analytic_C}
\mathcal{C}(t)=\left\{ 
\begin{aligned}
&\mathcal{B}_{1}(2\alpha t)/(\alpha t),\quad \delta=0;\\
&\exp(-\alpha^{2}t^{2}/2),\quad \delta=1/2;\\
&{\rm sech}(\alpha t),\quad \delta=1,
\end{aligned}
\right.
\end{align} 
where $\mathcal{B}_{1}$ denotes the Bessel function of the first kind. For generic $\delta$, there is no analytical solution but the equation can be easily solved numerically. It is found that non-monotonic oscillation exists when $\delta<0.5$ but disappears when $\delta>0.5$. 

\textit{Summary and Outlook.} In summary, we have found three universal properties of the autocorrelation functions with the help of the Krylov complexity. To reveal these universal properties, two key findings are the characteristic parameter $b^\prime[1]/b[1]$ and scaling time with $b[1]$. We emphasize that these universal properties are shared by different operators in different systems. On the theory side, our results bring out the generic connection between complexity and correlation in quantum many-body systems that deserves further theoretical investigations. On the experimental side, the quench experiments recently performed in NMR and cold atom systems directly measure the autocorrelation function \cite{D. Wei,Peng}, and our results can be straightforwardly verified in these experiments.    

\textit{Note Added.} Oscillatory versus non-oscillatory behavior of auto-correlation function in random spin model has been discussed in Ref. \cite{Zhou}.

\textit{Acknowledgement}. We thank Pengfei Zhang, Tian-Gang Zhou, Chang Liu, Yingfei Gu and Shang Liu for helpful discussions. This work is supported by NSFC 12174300 (RZ), the National Key R$\&$D Program of China 2018YFA0307601(RZ), Tang Scholar (RZ), Innovation Program for Quantum Science and Technology 2021ZD0302005 (HZ), the Beijing Outstanding Young Scholar Program (HZ) and the XPLORER Prize (HZ).

\end{document}